# Stochastic Optimization Based Study of Dimerization Kinetics

Srijeeta Talukder[a], Shrabani Sen[a], Ralf Metzler[b,c,*], Suman K Banik[d,*], Pinaki Chaudhury[a,*]

[a]Department of Chemistry, University of Calcutta, 92 A P C Road, Kolkata 700 009, India
[b]Institute for Physics & Astronomy, University of Potsdam, D-14476 Potsdam-Golm, Germany
[c]Physics Department, Tampere University of Technology, FI-33101 Tampere, Finland
[d]Department of Chemistry, Bose Institute, 93/1 A P C Road, Kolkata 700 009, India

**Abstract:**

We investigate the potential of numerical algorithms to decipher the kinetic parameters involved in multi-step chemical reactions. To this end we study a dimerization kinetics of protein as a model system. We follow the dimerization kinetics using a stochastic simulation algorithm and combine it with three different optimization techniques (Genetic Algorithm, Simulated Annealing and Parallel Tempering) to obtain the rate constants involved in each reaction step. We find good convergence of the numerical scheme to the rate constants of the process. We also perform a sensitivity test on the reaction kinetic parameters to see the relative effects of the parameters for the associated profile of the monomer/dimer distribution.

## 1. Introduction

The principles of chemical kinetics constitute one of the corner stones in the study of chemical and biological reaction networks. The evaluation of correct individual step based pathways (both the nature of the reaction as well as the correct magnitude of the rate constant) present in a multi-step reaction scheme is central in establishing a complete reaction model in any multi-step reaction process. The conventional way to study reaction kinetics is to write down the mean field rate equations for the process, integrate them and follow the variation in the concentration of each species involved as a function of time. However this strategy is too simplistic and fails in situations where the number of reacting species is small, since for small number of particles fluctuations in the species population become relevant. In the conventional rate equations approach, it is assumed that the process is both continuous and deterministic. However in reality, in particular in biochemical reactions in living cells, these assumptions often fail. Thus, reactants may occur at nanomolar rates. Reaction kinetics at such low concentrations are intrinsically discrete and stochastic. The stochastic simulation algorithm (SSA) is an elegant formulation to incorporate these effects and predict correct results in a complicated multi-step reaction network [1, 2]. Applications of SSA range from the study of simple schemes like two step consecutive reactions or parallel reactions [2], to biological systems like the dynamics of biopolymers such as DNA [3, 4]. The rate constants associated with each individual step in a multi-step reaction scheme might not always be known a priori, or there might be a range of values of the rate constants, for which predictions for the overall reaction are compatible with experimental data. The correct prediction of all individual rate constants is not always an easy task and involves an optimization process. If an optimization scheme can be linked to SSA, then it should be possible to evaluate a correct set of reaction parameters, quantifying the complete kinetic behavior of a reaction network. Here we analyze in detail the application of stochastic optimization schemes to the dimerization kinetics of proteins. It is also an experimentally well studied kinetics [5, 6].

*Corresponding author
*Email addresses:* rmetzler@uni-potsdam.de (Ralf Metzler),
skbanik@bic.boseinst.ernet.in (Suman K Banik),
pinakc@rediffmail.com (Pinaki Chaudhury)
*Telephone Number* : 91-9830480149 (Pinaki Chaudhury)

Optimization schemes can generally be classified into two categories, deterministic and stochastic optimization. The main difference between these two schemes is that deterministic ones are not truly global optimizers whereas the stochastic ones are. Stochastic optimizers are not gradient based and incorporate the principle of stochasticity to arrive at correct solutions. Here we use three techniques separately in conjunction with the SSA scheme to evaluate the correct set of reaction rate constants, these being Simulated Annealing (SA), Genetic Algorithm (GA) and Parallel Tempering (PT). These algorithms are also known as natural algorithms as they draw their working philosophy from natural processes. In particular, they are robust and able to find solutions to complex problems with consummate ease, unlike deterministic methods. Let us summarize the fundamental properties of the three stochastic optimizers:

(i) SA is a global optimization technique which mimics the process of annealing in metallurgy to design a mathematical optimization scheme. Thus the energy landscape of the search space is initially sampled at a high temperature, such that thermal fluctuations may easily lift the optimizer out of deeper minima. On decreasing the temperature the search is guided towards the global minimum. SA has been widely used in the last few decades with wide spread applications ranging from the solution of the travelling salesman problem, solving differential equations, finding structures of novel materials and studies involving structure and dynamics in quantum chemistry [7-15].

(ii) PT is a method based on replica exchange among randomly created configurations. PT has been successfully used to solve highly dimensional optimization problems with applications in the determination of structural features in proteins and other polymeric materials, the study of spin glasses and other solid state systems, or phase transitions in clusters formed by hydrogen bonding or other van-derWaals forces [16-23].

(iii) GA uses the concepts of genetics and mimics the natural process of selection, crossover and mutation as present in living systems. GA has been widely used, its applications ranging from solutions of differential equations, geometry optimization in large molecules and clusters, or the design of laser pulses to follow dissociation dynamics of molecular systems [24-33].

As we are interested in the evaluation of the correct kinetic rate parameters of a multi-step reaction, it is important to know a-priori the relative importance of each parameter. Sensitivity analysis is an important statistical technique which can give us an insight into this question. Such sensitivity analyses have been used widely to assess how individual parameters influence the dynamics when perturbed from their expected values [34, 35]. Such an analysis provides relevant information to decide the actual reaction scheme of some chemical kinetics. If a parameter is more sensitive than the other small deviations from the mean value will leave its mark on the kinetics by introducing large deviations from expected trends, while for insensitive parameters, the effect will be minimal. One can devise various measures for quantitatively calculating the sensitivity measures for various parameters like the Fourier amplitude based sensitivity test [36, 37], and others. This analysis leads to a more in-depth understanding of any kinetic scheme.

Here we follow the three combined schemes: GA+SSA, SA+SSA and PT+SSA to evaluate the correct set of rate constants necessary to explain all features in the reaction scheme of the dimerization of a protein. We perform a sensitivity test on the rate parameters in our model and use the findings to see if an optimization strategy influenced by the sensitivity test can lead to quicker convergence.

## 2. The dimerization model

In our study we have chosen a minimal kinetic scheme of protein dimerization [38], for which the minimal kinetic steps can be described by the following elementary reactions,

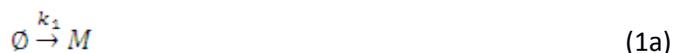

$$\emptyset \overset{k_1}{\rightarrow} M \tag{1a}$$

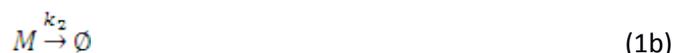

$$M \overset{k_2}{\rightarrow} \emptyset \tag{1b}$$

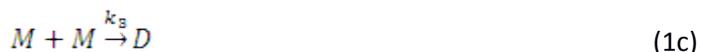

$$M + M \overset{k_3}{\rightarrow} D \tag{1c}$$

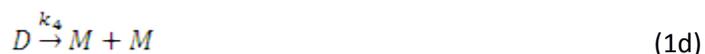

$$D \overset{k_4}{\rightarrow} M + M \tag{1d}$$

$$D \xrightarrow{k_5} \varnothing \qquad (1e)$$

where the $k_i$ are the rate constants for the individual reactions steps (1a) to (1e), and M and D denote the number of monomers and dimers of the protein, respectively. To keep the dynamics simple we neglect effects of cell growth and cell division in the model, i.e., we assume that the reaction occurs at fixed volume.

The marginal probability distribution function for monomer and dimer can be defined as

$$\tilde{P}_x(y,t) = \int P(x,y,t)dx \qquad (2a)$$

$$\tilde{P}_y(x,t) = \int P(x,y,t)dy \qquad (2b)$$

where x and y are the number of monomer and dimer molecules at time t, respectively. As the system deals with small numbers of molecules the time evaluation should be obtained by using stochastic formulation, i.e. by solving master equation. But it is often quite tedious to solve such a master equation analytically for complicated systems like the present one. Consequently one has to resort to numerical methods to quantify the underlying process, for instance by using SSA.

Typically, in SSA a probability density function $P(\tau,\mu)d\tau$ is introduced for a given state at time $t$, which is a measure for the probability that within the infinitesimal time interval $(t + \tau, t + \tau + d\tau)$ the $\mu$th reaction will occur, where the index $\mu$ stands for a given reaction step. $P(\tau, \mu)d\tau$ is supposed to follow Poissonian statistics such that

$$P(\tau,\mu) = a_\mu exp - (a_0\tau) \qquad (3)$$

Where

$$a_\mu = h_\mu c_\mu, a_0 = \sum_{\mu=1}^{N} h_\mu c_\mu \qquad (4)$$

In the latter relation $h_\mu$ is the number of molecules of the species involved in the $\mu$th reaction step, $c_\mu$ is the respective rate constant and N is the number of reaction channel. The time step $\tau$ is defined via

$$\tau = \frac{1}{a_0} ln \frac{1}{r_1} \qquad (5)$$

In SSA, the underlying randomness is introduced in the choice of and $\mu$ as follows: $\mu$ be the integer for which

$$\sum_{\nu=1}^{\mu-1} a_\nu < r_2 a_0 \le \sum_{\nu=1}^{\mu} a_\nu \qquad (6)$$

where $r_1$ and $r_2$ are random numbers between 0 to $1^{1,2}$. $\tau$, in Eq (5), thus can never be negative as $r_1 \leq 1$ (the derivation of Eq (5) is given in Appendix). Utilizing SSA in the present reaction scheme given by Eqs. (1a-1e), one may calculate the equilibrium marginal distribution profile separately for monomer and dimer.

## 3. Stochastic Optimization

We employ the three optimization techniques SA, GA, and PT to determine the optimum set of kinetic parameters in our kinetic scheme. These optimization techniques are not gradient based and use stochastic principles, hence are known as stochastic optimizers. All the simulations are started with an arbitrary parameter set which is obtained by perturbing the literature value of the rate constants in the dimerization of protein to a preset extent. The same initial parameter set has been taken for all the three optimization schemes. To follow the progress of the optimization, we compare the probability distribution profile of the monomers and dimers of the protein obtained by using the literature value of the rate constants in the SSA with the distribution profile for various sets of the rate constants obtained from different iterations of the simulations. The parameters are said to be optimized if the distribution profile for a set of parameters coincides with the distribution profile from literature within 0.01%.

During the simulation, the parameter set of rate constants obtained in each iteration is fed into the SSA to produce the distribution profile corresponding to this output of the optimizer. Then, an objective function, popularly known as the cost function, is calculated to measure the extent of the difference between the present distribution profile and the profile obtained for the literature value of the parameter set (expected distribution profile). The cost function is basically the cumulative differences in probabilities for the two distribution profiles, for different species (in our system for the monomer and the dimer only). While calculating the difference in the two distributions, the magnitudes of the two functions at certain discrete points have been taken. The index 'i' in Eq.(7) refers to the discrete points, at which the values of the two functions have beencompared.

$$cost = \sum_{m=1}^{n} \sum_{i=1}^{k} \left( \tilde{P}_i(m_i) - \tilde{P}(m_i) \right)^2 \tag{7}$$

where $\tilde{P}_i(m_i)$ and $\tilde{P}(m_i)$ are the probabilities of the $m^{th}$ species at the $i^{th}$ grid point (whereas n is the total number of species and k is the total number of grid points) in the distribution profile obtained by feeding the literature value of the rate constants [38] in SSA and the profile for the rate parameters at a optimization step respectively. The cost has to be minimized with simulation and for the optimum solution it must tend to zero. Fig. 1 shows the distribution profiles for literature value and for a set of optimized parameters, which actually coincides with the objective distribution profile.

Assigning of cost function utterly depends on the specificity of the problem. The above equation may not be the ultimate way to define cost function. One may define cost with respect to the time series profile of mean and variance also. Then the objective becomes to reach the mean and variance profile for the literature value, in course of optimization.

$$cost = \sum_{m=1}^{n} \sum_{i}^{k_t} \left( [m]_i^l - [m]_i \right)^2 + \left( Var(m)_i^l - Var(m)_i \right)^2 \qquad (8)$$

$[m]_i$ and $Var(m)_i$ are the concentration and the variance of $m^{th}$ species at $i^{th}$ time respectively. Term with notation *l* is for the expected profile and the other is obtained from simulation. In this case also the cost would be theoretically zero for the optimal solution. Fig. 2 represents such profile for literature value of rate constants and the profile for the set of rate constant acquired from an optimization run.

## 4. Sensitivity Analysis

Generally, in any chemical or biochemical network not all the parameters hold equal priority. A sensitivity analysis is conducted to determine which input parameters contribute the most to the output variable, which parameters are insignificant, whether the input parameters do interact among themselves, whether the interaction is physically explainable, and, after all, to search for the optimal regions within the parameters space for use in a subsequent calibration study. One can say a system is sensitive with respect to a parameter if a small change to this parameter affects the output abruptly.

For a quantitative estimation of the sensitivity of the rate parameters, we use a variance based sensitivity analysis test. The idea of this analysis is taken from Saltelli et al. [34]. They report a comparative discussion of different sensitivity analysis techniques in order to reduce the computational cost of running the model. The main idea was developed by Cukier and coworkers [36, 37] in the 1970-ies, and was known as Fourier amplitude sensitivity test (FAST). In the present work we adopt the implementation of the FAST based sensitivity test as used by Saltelli et al. [34].

The variance in output with input parameter set having one parameter fixed at some value is defined by the term $V_{X_{-i}}(Y|X_i=X_{-i})$, where Y is the output factor and the subscript $X_{-i}$ of V is to mention that the variance is taken over all other input parameters other than $X_i$ which is fixed at $X_{-i}$ . This is generally less than the variance with fully random input set V(Y), but may depend on the magnitude of the fixed parameter. To remove this type of parameter dependence an average of the variance over the different values of the fixed parameter is estimated by $E_{X_i}(V_{X_{-i}}(Y|X_i=X_{-i}))$. We may write the total variance V(Y) as follows [34]:

$$V(Y) = E_{X_i}\left(V_{X_{-i}}(Y|X_i)\right) + V_{X_i}\left(E_{X_{-i}}(Y|X_i)\right) \tag{9}$$

Thus a FAST-based sensitivity index is defined simply as

$$S_i = \frac{V_{X_i}\left(E_{X_{-i}}(Y|X_i)\right)}{V(Y)} \tag{10}$$

Using the above mentioned idea we discuss the sensitivity of the model parameters in the next section.

## 5. Results and Discussion

Our main focus is to evaluate the optimal set of kinetic parameters for the scheme given by Eqs. (1a-1e) of protein dimerization. All stochastic optimizers: SA, GA, and PT, turned out to decipher the optimal set of kinetic parameters. The rate constants obtained from simulations are in good proximity to the literature value. The optimized parameter values (average of the five runs) in each simulation procedure, as well as the literature values[38] are shown in Table 1.

We followed the approach to the converged results for the five kinetic parameters of the protein dimerization model in the three different schemes. The results are shown in Fig. 3. For each technique we show five simulation runs, as shown in the graph. The literature value of each parameter is shown as black dashed line. The parameters $k_1$ and $k_2$ show a good convergence within a very short range around the literature value, but the other rate parameters show a spread, for the reason see the discussion below. It is also evident from Fig. 3 that GA and PT runs take fewer steps (about 60 to 70) to converge to reasonably convincing solutions, while SA takes about 100 steps. But in terms of computational time required, GA seems to be most efficient method followed by SA and PT. This is expected on theoretical lines, as GA and PT process a number of trial solutions simultaneously, while SA improves on a single starting solution.

Let us now come to the Sensitivity Analysis to classify which rate constants are the most delicate. We apply a fixed perturbation (5%) onto a given rate constants at a time, keeping the other rate parameters unperturbed with respect to the literature value. We then use SSA to return the equilibrium probability distribution for the perturbed run. By comparing the discussed distribution function with the theoretical one, we qualitatively judge the sensitivity of the rate parameters (plots shown in Figs. 4). In our case the dimerization kinetics is more sensitive to $k_1$ and $k_2$ than the others. This result justifies the observation in Fig. 3. The more sensitive parameters should converge to a small range around the target value whereas relatively less sensitive ones exhibit a wider spread at the end of the optimization.

The sensitivity index ($S_{k_i}$) for the output monomer and dimer concentrations are calculated separately with respect to each input parameter at different time for the protein dimerization kinetics. Fig. 5 depicts the plot of sensitivity index against time. Higher sensitivity for a particular parameter indicates that the system becomes more sensitive to that parameter than the others. Our results depict that both the monomer and dimer concentrations are more sensitive with respect to $k_1$ and $k_2$ than to $k_3$, $k_4$, and $k_5$. Fig. 5 also shows that as a function of time, $S_{k_1}$ decreases and $S_{k_2}$ increases, and a crossover of sensitivity occurs. This crossover reflects the physical idea that initially the system becomes sensitive to the rate constant, which produces monomers from a source. As time progresses, due to adequacy of monomer concentrations, the sensitivity index shifts to the reverse rate constant of the reaction. This type of crossover is also observed on a smaller scale in $k_3$ and $k_4$.

Since $k_1$ and $k_2$ are reflected to be the more sensitive parameters, they are needed to be explored more than the rest of the rate parameters. This idea is incorporated during the parameter evolution by using stochastic optimization techniques. If we assign higher probability to the sensitive parameters, to be sampled than the others, convergence occurs rapidly during optimization. Fig. 6 clearly shows that on applying 80% weight on $k_1$ and $k_2$ to be sampled, (20% weight on $k_3$, $k_4$ and $k_5$) the cost function falls more rapidly than the sampling with 60% weight on $k_1$ and $k_2$ (40% weight on $k_3$, $k_4$ and $k_5$). This in turn is obviously faster than a run, in which equal weight is assigned for sampling of each rate parameter. The trends are in a similar line for runs involving SA, GA and PT. This strategy of doing biased optimization runs, involving higher weight to sample more sensitive parameters, will certainly contribute to the decrease in computational cost.

It is also important as a concluding remark in this section is to have an idea of the robustness of the three used numerical algorithms in deciphering the rate constants. In the present study we have selected the initial trail set of rate constants from a Gaussian distribution, whose peak corresponds to the literature value and a half width of 10% of the respective rate constant. This is a moderately large perturbation. The convergence from this initial set is quite close to the values reported in literature. As we have also done a sensitivity analysis on each of the five calculated reaction rate parameters, a check can also be made on the relative importance/rigidity that an evaluated value can have. Less is the sensitivity of a particular data, a greater spread from the reported value can happen even while matching the correct dimer-monomer distribution profile.

## 6. Conclusions

We have shown that stochastic optimization techniques in conjunction with SSA can help determining kinetic parameters in multi-step kinetic schemes. All the three optimizers SA, GA and PT perform equally well to predict the values of the rate constants. We have also shown that an optimization study guided by findings from Sensitivity Analysis can help us distinguish between the parameters based on its importance and if these are incorporated into the optimization, a quicker convergence can be achieved. This strategy of initially doing a sensitivity analysis and segregating the rate parameters into zones of importance and then subsequently using stochastic optimization techniques to decipher them can be an important strategy for studying kinetics in complex biochemical networks, where the numbers of rate parameters can be numerous. The number of ordinary differential equations to be solved in such cases are also very large. An unbiased strategy of allocating equal samplings to each and every rate parameter will make the process of finding the solution, computationally costly and tedious. A stochastic search in conjunction with the sensitivity analysis will be much more efficient.

**Acknowledgments**

ST acknowledges the financial support form UGC, New Delhi, for granting a Junior Research Fellowship. SS thanks the UGC, New Delhi for granting a D.S. Kothari post-doctoral fellowship. RM acknowledges funding through the Academy of Finland's FiDiPro scheme. PC wishes to thank The Centre for Research on Nano Science and Nano Technology, University of Calcutta for a research grant [Conv/002/Nano RAC (2008)]. SKB acknowledges support from Bose Institute through Institutional Programme VI - Development of Systems Biology.

## Appendix

To generate a real random number x, which follows a probability density function P(x), one has to consider another function F(x)

$$F(x) = \int\limits_{-\infty}^{x} P(x')dx'$$

F(x) is the probability distribution function which satisfy

$$F(x) = r$$

Where $r$ is basically a random number from the uniform distribution between 0 to 1

Then,

$$x = F^{-1}(r)$$

In Eq (5) τ (a real random number) follows the probability density function P(τ) [1,2].

$$P(\tau) = a_0 \exp(-a_0 \tau) \qquad \text{for } 0 \leq \tau \leq \infty$$

$$= 0 \qquad \text{elsewhere}$$

Now,

$$F(\tau) = \int_0^\tau P(\tau') d\tau'$$

$$r = 1 - \exp(-a_0 \tau)$$

(1-r) is also a random number, thus τ becomes

$$\tau = \frac{1}{a_0} ln \frac{1}{r}$$

**Figure Caption**

Figure 1: Plot of distribution profile of monomer and dimer. The solid line denotes the profile for literature value of rate parameters and the open circles represents the profile for optimized set. Red and blue colors are for depicting monomer and dimer respectively.

Figure 2: Plot of mean and variance of monomer and dimer concentration with time. Panel (a) is the time series of mean and panel (b) is that of variance. The solid line denotes the profile for literature value of rate parameters and the open circles represents the profile for optimized set. Red and blue colors are for depicting monomer and dimer respectively.

Figure 3: (color online) Kinetic parameters ($k_1$, $k_2$, $k_3$, $k_4$, and $k_5$) versus number of SA, GA and PT steps. The black dashed line represents the values of the kinetic parameters taken from Adalsteinsson *et al.* [38], while the step like lines (red, blue, green, cyan, and magenta) are the results of five different SA, GA, and PT runs.

Figure 4: (color online) Expected probability distribution (solid line) and distribution obtained from SSA by perturbing one parameter at a time (dashed lines). Red and blue represent monomer and dimer distributions, respectively. In panels (a) to (e) the perturbed parameters were, respectively, $k_1$, $k_2$, $k_3$, $k_4$ and $k_5$.

Figure 5: (color online) Sensitivity index, Ski for i = 1 − 5 versus time for monomers and dimers. The red, green, blue, magenta, and cyan colored lines are for $k_1$, $k_2$, $k_3$, $k_4$, $k_5$, respectively.

Figure 6:(color online) Optimization profile against SA, GA, and PT steps (in log scale). Red line: Simulation with equal weight to all the kinetic parameters. Green line: biased Simulation to 60 % weight to $k_1$ and $k_2$, Blue line: biased Simulation to 80 % weight to $k_1$ and $k_2$.

**Table 1: Comparison of kinetic parameter values. Units of $k_1$, $k_2$, $k_3$, $k_4$ and $k_5$ are nM min$^{-1}$, min$^{-1}$, nM$^{-1}$ min$^{-1}$, min$^{-1}$ and min$^{-1}$, respectively.**

| Parameter | Literature[38] | Simulation | | |
| --- | --- | --- | --- | --- |
| | | SA | GA | PT |
| $k_1$ | 50.0 | 49.42 | 49.41 | 49.44 |
| $k_2$ | 1.02 | 1.007 | 1.008 | 1.009 |
| $k_3$ | 0.01 | 0.009 | 0.009 | 0.009 |
| $k_4$ | 0.1 | 0.10 | 0.099 | 0.098 |
| $k_5$ | 0.2 | 0.019 | 0.019 | 0.019 |

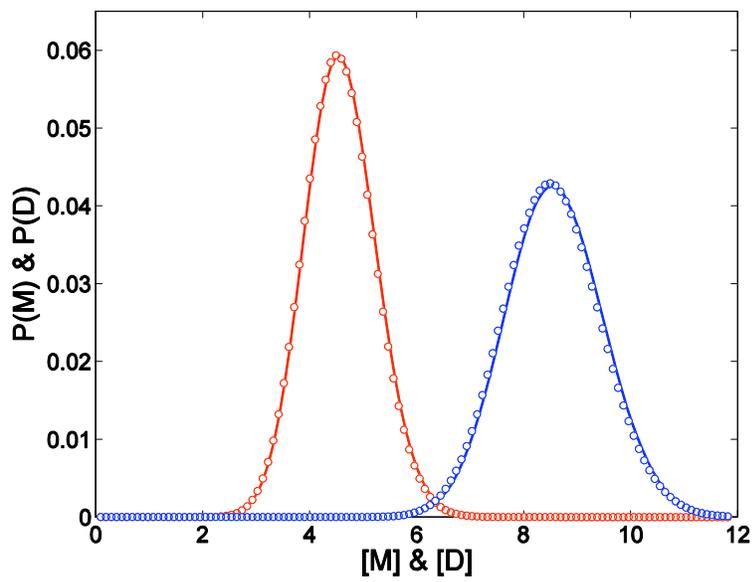

Figure 1

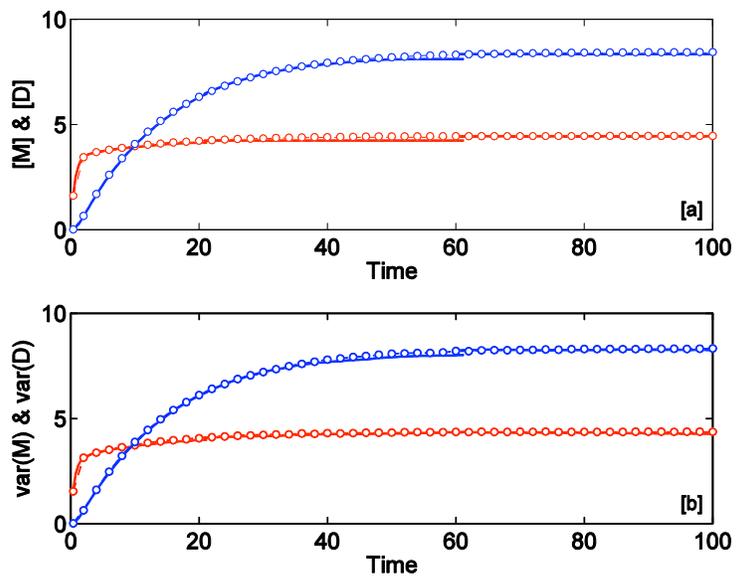

Figure 2

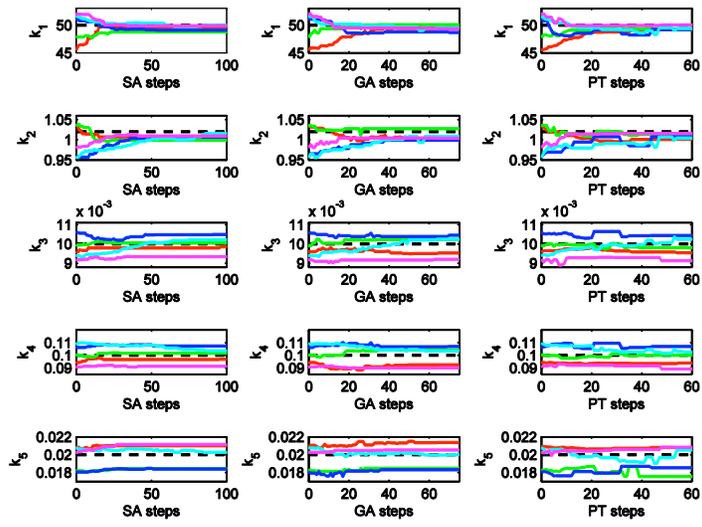

Figure 3

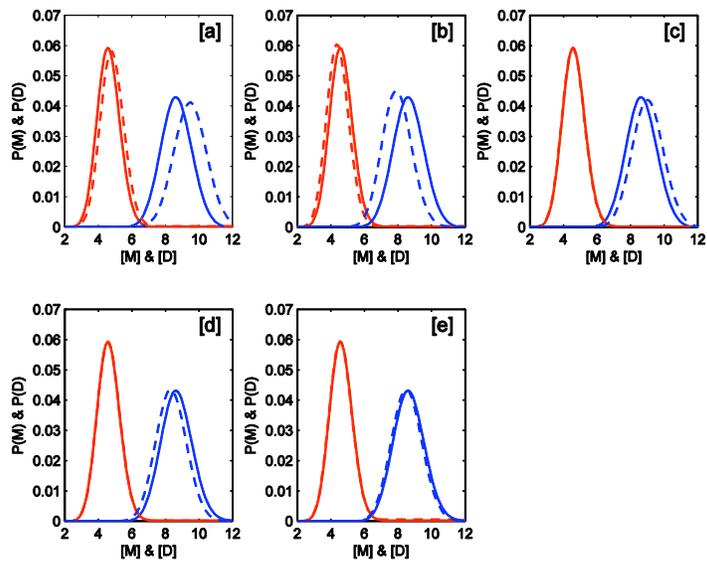

Figure 4

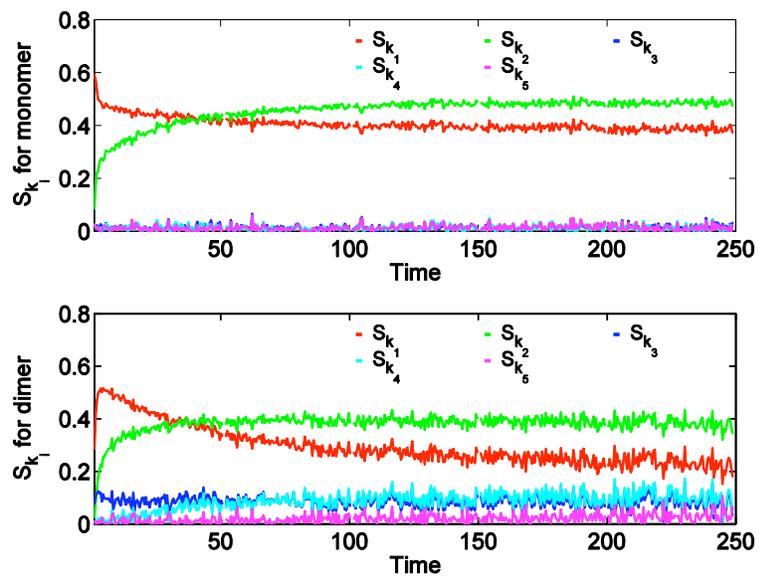

Figure 5

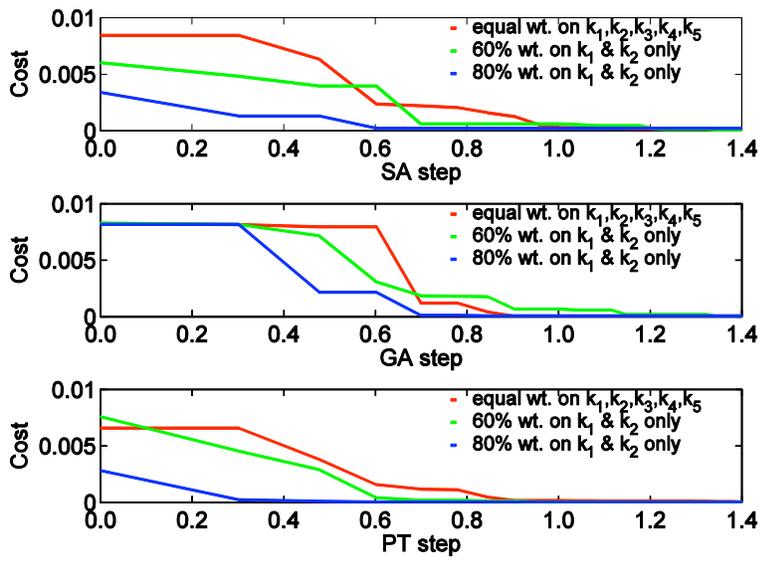

Figure 6